\documentclass[%
 reprint,
superscriptaddress,
nofootinbib,
 amsmath,amssymb,
aps, prl
longbibliography
]{revtex4-2}

\usepackage{graphicx}
\usepackage{dcolumn}
\usepackage{bm}
\usepackage{physics}
\usepackage{bbm}

\begin{document}

\preprint{APS/123-QED}

\title{Variance extrapolation method for neural-network variational Monte Carlo}

\author{Weizhong Fu}
\affiliation{ByteDance Research, Zhonghang Plaza, No. 43,  North 3rd Ring West Road, Haidian District, Beijing, People’s Republic of China}
\affiliation{School of Physics, Peking University, Beijing 100871, People’s Republic of China}

\author{Weiluo Ren}
\email{renweiluo@bytedance.com}
\affiliation{ByteDance Research, Zhonghang Plaza, No. 43,  North 3rd Ring West Road, Haidian District, Beijing, People’s Republic of China}

\author{Ji Chen}
\email{ji.chen@pku.edu.cn}
\affiliation{School of Physics, Peking University, Beijing 100871, People’s Republic of China}
\affiliation{Interdisciplinary Institute of Light-Element Quantum Materials, Frontiers
Science Center for Nano-Optoelectronics, Peking University, Beijing 100871, People’s Republic of China}

\date{\today}

\begin{abstract}
Constructing more expressive ansatz has been a primary focus for quantum Monte Carlo, aimed at more accurate \textit{ab initio} calculations.
%
However, with more powerful ansatz, e.g. various recent developed models based on neural-network architectures, the training becomes more difficult and expensive, which may have a counterproductive effect on the accuracy of calculation.
In this work, we propose to make use of the training data to perform variance extrapolation when using neural-network ansatz in variational Monte Carlo.
We show that this approach can speed up the convergence and surpass the ansatz limitation to obtain an improved estimation of the energy.
Moreover, variance extrapolation greatly enhances the error cancellation capability, resulting in significantly improved relative energy outcomes, which are the keys to chemistry and physics problems.
\end{abstract}

\maketitle


\section{Introduction}
Recent decades have brought significant advances in simulating complicated physical systems using computers, making it possible to model and understand intricate phenomena with unprecedented accuracy and detail.
Consequently, computer simulation is now considered the ``third way'' of doing science, bridging the gap between theory and experiment~\cite{foulkes2001quantum}.
In the fields of physics and chemistry, a crucial problem lies in determining the electronic structure through solving the Schr\"{o}dinger equation, a fundamental task in the realm of quantum mechanics~\cite{pople1999nobel, kohn1999nobel}.
%
%
%
Although the underlying physical laws have been well-understood for almost a century, ``the difficulty is only that the exact application of these laws leads to equations much too complicated to be soluble.'', as Paul Dirac famously wrote~\cite{dirac1929quantum}.
Solving the 3$N$-dimensional Schr\"{o}dinger equation describing a system containing $N$ interacting electrons seems impossible, but this is exactly what quantum Monte Carlo (QMC) methods allow us to do~\cite{foulkes2001quantum}.

QMC comprises of a wide range of stochastic methods based on random sampling~\cite{austin2012quantum}.
One of the most frequently used methods is the variational Monte Carlo (VMC)~\cite{mcmillan1965ground, ceperley1977monte}.
The basic idea of VMC involves assuming a particular wavefunction ansatz and applying the variational principle to optimize it towards the ground state.
%
%
Apparently, the accuracy of VMC is restricted by the expressiveness of the ansatz, which motivates the construction of more powerful ansatz.
%
%
Recently, various neural-network-like wavefunction ansatz have been proposed and greatly improved the accuracy of VMC~\cite{carleo2017solving, schutt2017schnet, gao2017efficient, cai2018approximating, cai2018approximating, han2019solving, luo2019backflow, pfau2020, choo2020fermionic, spencer2020better, hermann2020deep, von2022self, lin2023explicitly, abrahamsen2022taming, gerard2022gold, pescia2022neural, wilson2023neural, scherbela2022solving, li2022, gao2021ab, gao2022sampling, gao2023generalizing, scherbela2023towards, barrett2022autoregressive, zhao2023scalable}.
For convenience, we refer to this as neural-network VMC (NN-VMC) in the following text.
%
FermiNet~\cite{pfau2020, spencer2020better} and DeepSolid~\cite{li2022} are examples of successful wavefunction ansatz for molecular and periodic systems respectively.

Despite the success of NN-VMC in accuracy, its widespread use is limited by the high computational cost incurred by the large number of parameters to be optimized and the long training process required for convergence. 
This results in the unsatisfactory error cancellation capability~\cite{ren2023towards}, which is essential for estimating relative energy, such as ionization energy, dissociation energy and so on.
Previously when calculating relative energies, identical hyperparameters were shared across different systems, and the discrepancies between total energy outcomes were regarded as the results.
This strategy, however, is fair but unbalanced, since a relative energy involves the energies of two different systems, but a more complicated system may require a larger network to handle, and the speed of convergence can differ across different systems.
%

In this work, we propose to make use of the large amount of energy-versus-variance data obtained during the prolonged training process to perform zero variance extrapolation in NN-VMC.
We evaluate this extrapolation approach across a range of systems, including N$_2$ molecules of varying bond lengths and some well known periodic systems, using respectively FermiNet and DeepSolid as the wavefunction ansatz.
We also incorporate the variance matching method for N$_2$ molecules.
Although the results of variance matching method appear to be unsatisfactory, our proposed extrapolation approach demonstrates high performance across all systems.
Notably, the extrapolation is solely based on training data, without incurring additional computational costs.
In addition to obtaining better relative energy, it can surpass the limitation of the wavefunction ansatz and mitigate the bias caused by the inadequate training convergence, thus also delivering an better evaluation of total energy, which is also discussed.

\section{Methods}
\subsection{VMC theory}
VMC is a stochastic method for solving quantum many-body problems. For a complex quantum system, the many-body Hamiltonian operator can be expressed as
\begin{equation}
\begin{split}
    \hat{H} =  - \frac{1}{2}\sum_{i}\nabla^{2}_{i} - \sum_{I, i}\frac{Z_{I}}{\left|\mathrm{\mathbf{r}}_{i}-\mathrm{\mathbf{R}}_{I}\right|} \\ +  \sum_{i<j}\frac{1}{\left|\mathrm{\mathbf{r}}_{i}-\mathrm{\mathbf{r}}_{j}\right|} + \sum_{I<J}\frac{Z_I Z_J}{\left|\mathrm{\mathbf{R}}_{I}-\mathrm{\mathbf{R}}_{J}\right|}, 
\end{split}
\end{equation}
where $i, j$ are subscripts for electrons, and $I, J$ for nuclei. $Z_I$ denotes the charges, and $r_i, R_I$ denote the positions.
Under Born-Oppenheimer approximation~\cite{born1927quantentheorie}, the main goal is to obtain the ground state wavefunction and corresponding energy of the stationary Schr\"{o}dinger equation,
\begin{equation}
    \hat{H}\ket{\psi_i} = E_i\ket{\psi_i}, i=0,1,2...
\end{equation}
To accomplish this, VMC employs Monte Carlo methods to handle the high dimensional energy integral and variational principles to optimize the wavefunction towards the ground state. 
Given a wavefunction ansatz $\psi_\theta(\mathbf{r})$, its energy expectation is defined as
\begin{equation}
    \begin{split}
        E\left[\psi_\theta\right] & = \frac{\bra*{\psi_\theta}\hat{H}\ket*{\psi_\theta}}{\braket{\psi_\theta}} = \frac{\int\psi^*_\theta(\mathbf{r})\hat{H}\psi_\theta(\mathbf{r})\mathrm{d}\mathbf{r}}{\int\psi^*_\theta(\mathbf{r})\psi_\theta(\mathbf{r})\mathrm{d}\mathbf{r}} \\
        & = \int p(\mathbf{r}) E_L(\mathbf{r}) \mathrm{d}\mathbf{r} = \mathbb{E}_{p(\mathbf{r})}\left[E_L\right],
    \end{split}
\end{equation}
where $p(\mathbf{r}) = \frac{\left|\psi_\theta(\mathbf{r})\right|^2}{\int \left|\psi_\theta(\mathbf{r})\right|^2 \mathrm{d}\mathbf{r}}$ is the normalized wavefunction squared, i.e. the probability density function, and $E_L(\mathbf{r})=\frac{\hat{H}\psi_\theta(\mathbf{r})}{\psi_\theta(\mathbf{r})}$ is the so-called local energy function.
The outcome of this energy integral must be greater than the ground state energy, which is the target of VMC, so we just need to optimize the wavefunction towards lower energy,
\begin{equation}
    -\nabla_\theta E\left[\psi_\theta\right]\propto \mathbb{E}_{p(\mathbf{r})}\left[\left(\mathbb{E}_{p(\mathbf{r})}\left[E_L\right] -E_L\right)\nabla_\theta \log \left|\psi_\theta \right|\right].
\end{equation}
Keep optimizing until convergence, and we will get the approximation of ground state wavefunction and corresponding energy, whose accuracy depends on the expressiveness of the wavefunction ansatz.
\begin{figure*}[!htbp]
    \includegraphics[width=160mm]{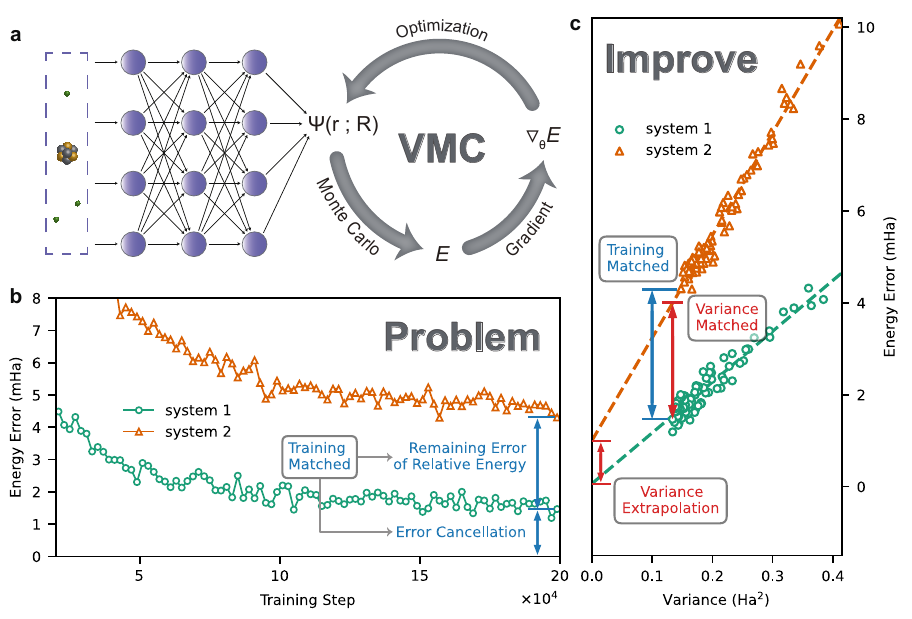}
    \caption{\label{schematic_sketch}\textbf{a}, A schematic workflow of NN-VMC. \textbf{b}, An existing problem of NN-VMC on relative energy calculation. The two curves show the process of energy error converging with training of two different systems.
    The curves are real calculations on N$_2$ molecule at two bond lengths, which will be described with more details in Sec. III.A. 
    The usual way to obtain the relative energy of these two systems is to compare the energy results at the same training step, which is labeled as ``Training Matched'' (TM). Since the y-coordinate represents the energy error here, the lower energy error equals the cancelled error, and the difference equals the remaining error of the relative energy. \textbf{c}, Two better ways to obtain relative energies, ``Variance Matched'' (VM) and ``Variance Extrapolation'' (VE). Along the training process, the energy and energy variance present a linear relationship, and the two dashed lines show the linear fitting results. The three double-headed arrows show the remaining errors of relative energy with three different methods. The variance matched result is obtained by comparing the energies at the same variance. The variance extrapolation result is obtained by comparing the y-intercepts of the two linear fitting curves. The training matched result is marked for comparison.}
\end{figure*}
\subsection{Better relative energy in NN-VMC}
NN-VMC reaps the benefit of a vast number of parameters to obtain high accuracy.
However, this comes at the cost of a lengthy optimization process, which unfortunately results in less satisfactory error cancellation capability when calculating the relative energy.
Since there is no established criterion for measuring the degree of convergence during the training, the common approach is to use the same optimization iteration in all the systems for calculating relative energies, as shown in Fig.~\ref{schematic_sketch}b and labeled as ``Training Matched'' (TM).
Nevertheless, this can lead to an unbalanced treatment if the optimization is inadequate for convergence because of the different speed of convergence in different systems.
Even if achieving complete convergence, this simple fairly-treating approach is still unbalanced as the more complicated system requires a larger network to obtain the same accuracy.

To achieve balanced treatments, matching a quantity related to the error in a wavefunction is a potential strategy, which has been utilized in various studies to obtain excitation energy~\cite{robinson2017a, pinedaflores2019, otis2020, garner2020variational, entwistle2023}.
The energy variance is selected as the matching quantity and is defined as:
\begin{equation}
\sigma_\theta^2=\frac{\bra{\psi_\theta } (\hat{H} - E)^2  \ket{\psi_\theta}}{\braket{\psi_\theta}},
\end{equation}
where $E = \langle\hat{H}\rangle_{\psi_\theta}$ denotes the energy expectation.
As shown in Fig.~\ref{schematic_sketch}c, the ``Variance Matched'' (VM) method is thus comparing the energies of two systems where they have the same variance.
To clarify further, we choose the data point with the most training steps from ``system 1'' and compare its energy result with ``system 2'', whose energy result is determined via linear regression at the corresponding variance.

Another strategy is the zero-variance extrapolation method, where the energy outcomes are extrapolated until the variance reaches zero as the real ground state should possess zero energy variance.
In VMC, when the wavefunction is optimized to be near the ground state, the energy and variance exhibit a linear relationship, as demonstrated in Fig.~\ref{schematic_sketch}c.
Although the fundamental reason for the linear relationship is not fully clear~\cite{kwon1993,kwon1998},
it leads to empirical extrapolation schemes that have been successfully used in various systems, including Fermi liquids and the Hubbard model~\cite{taddei2015, robledomoreno2022, hu2013, iqbal2016}.
The most commonly used schemes for extrapolation in previous studies are: 1) based on the converged results of different ansatz, such as different forms of backflow or different widths of the network~\cite{taddei2015, robledomoreno2022}, and 2) based on the results of different converged degrees with the same ansatz, such as different Lanczos steps~\cite{hu2013, iqbal2016}.
In other words, there are two different routes to obtain a range of wavefunctions with different ``distance'' to the ground truth. 
In this work, we employ extrapolation based on the training data at different converged levels of NN-VMC, as shown in Fig.~\ref{schematic_sketch}c, labeled as ``Variance Extrapolation'' (VE).

To obtain effective information from the noisy training data, a rolling window is employed to calculate the robust mean of the energy and variance, which means in each window the outlier data are masked.
%
%
We assume that the wavefunction in a window does not change much, and the energy data follow the normal distribution.
For a window containing $n$ data points, the data points whose absolute deviation of the energy from the median is larger than 3$\sigma$ are considered to be outliers, where $\sigma$ is the standard deviation.
In all the calculations we mentioned in this paper, we set the window size $n$ to be 2000.
After masking, the energy and variance outcomes of this window are calculated following:
\begin{equation}
\left\{
\begin{aligned}
    E_{out} & = \overline{e} \\
    V_{out} & = \overline{v} + \overline{e^2} - \overline{e}^2
\end{aligned}
\right. 
\end{equation}
where $e$ and $v$ denote the energy and variance of the remaining data after masking, and the overline denotes taking the average.
It is worth pointing out that in the training process of NN-VMC, there may be extreme outliers that can significantly impact the extrapolation outcome, thus making the masking step indispensable.

For more technical details about the selection of linear segments for extrapolation and the appropriateness of using training data rather than inference data, refer to Supplementary Note~1.

\subsection{Linear relationship in VE method}
In the VE method, we assume that there exists linear relationship between the energy and the energy variance when the wavefunction is nearing the ground state.
As mentioned earlier, the fundamental reason for the linear relationship remains unclear.
However, we can provide a possible explanation by offering a sufficient but unnecessary condition for such a relationship, following the idea of Kashima et al.~\cite{kashima2001path}.

%
In the late training period, assume that the neural-network state can be represented as a linear combination of ground state component $\ket{\psi_0}$ and another fixed eigenstate of Hamiltonian $\ket{\psi'}$. 
That is,
\begin{equation}
    \ket{\psi_\theta} = \alpha\ket{\psi_0}+\beta\ket{\psi'}.
\end{equation}
Here $|\alpha| \gg |\beta|$, corresponding to the ground state's dominance as the training process approaches convergence.
Without loss of generality, here we assume the normalization condition $\alpha^2+\beta^2 = 1$.
The energy expectation result is
\begin{equation}
    E=\langle\hat{H}\rangle_{\psi_\theta}=E_0+\beta^2(E'-E_0),
\end{equation}
where $E_0$ and $E'$ are the eigenvalues, i.e. the energies corresponding to $\ket{\psi_0}$ and $\ket{\psi'}$.
The energy variance result is
\begin{equation}
    V=\langle(\hat{H}-E)^2\rangle_{\psi_\theta}=\alpha^2\beta^2(E'-E_0)^2.
\end{equation}
We can thus derive the linear relationship
\begin{equation}
    E=kV+E_0,
\end{equation}
where the slope factor $k\approx\frac{1}{E'-E_0}$.


\section{Results and Discussion}
\subsection{N$_2$ molecule}

\begin{figure*}[!htbp]
    \includegraphics[width=160mm]{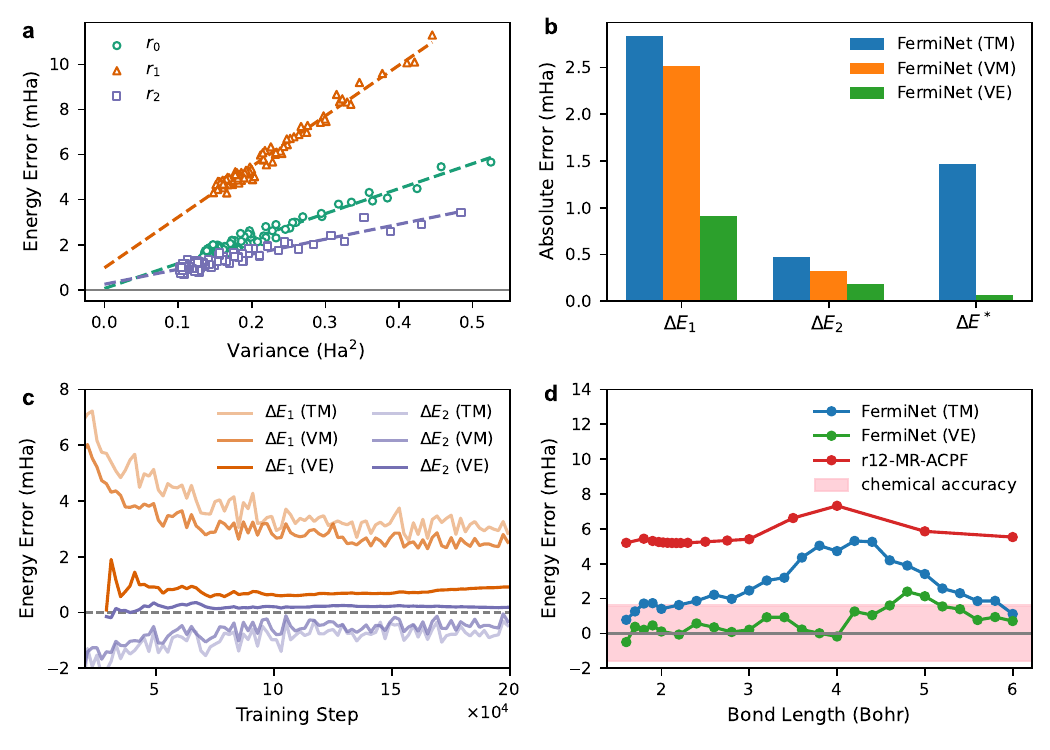}
    \caption{\label{var_matched_extrapolation}Effects of VM and VE methods on the N$_2$ molecule using FermiNet~\cite{pfau2020, spencer2020better} as the wavefunction ansatz. \textbf{a}, The linear relationship of the energy and energy variance results at three different bond lengths, which are respectively $r_0 = 2.0743$ Bohr at the equilibrium location, $r_1 = 4.0$ Bohr in the strongly correlated region and $r_2 = 10.0$ Bohr in the completely dissociated region. The dashed lines show the linear fitting results. \textbf{b}, The absolute errors of relative energies $\Delta E_1 = E(r_1) - E(r_0)$ and $\Delta E_2 = E(r_2) - E(r_0)$, with respectively TM, VM and VE methods. Additionally, we present the results of dissociation energy $\Delta E^*$ calculated following Eq.~\ref{eq6}. \textbf{c}, The errors of relative energies $\Delta E_1$ and $\Delta E_2$ with respect to the training steps, utilizing respectively TM, VM and VE methods. \textbf{d}, Results of TM and VE methods on N$_2$ dissociation curve. For comparison, the red line shows the state-of-the-art r12-MR-ACPF results under a modified basis set based on aug-cc-pV5Z~\cite{gdanitz1998accurately}. The pink area indicates the range of chemical accuracy. All the benchmarks in this figure are obtained from an experimental fitted curve~\cite{le2006accurate}.}
\end{figure*}

\begin{figure*}
    \includegraphics[width=160mm]{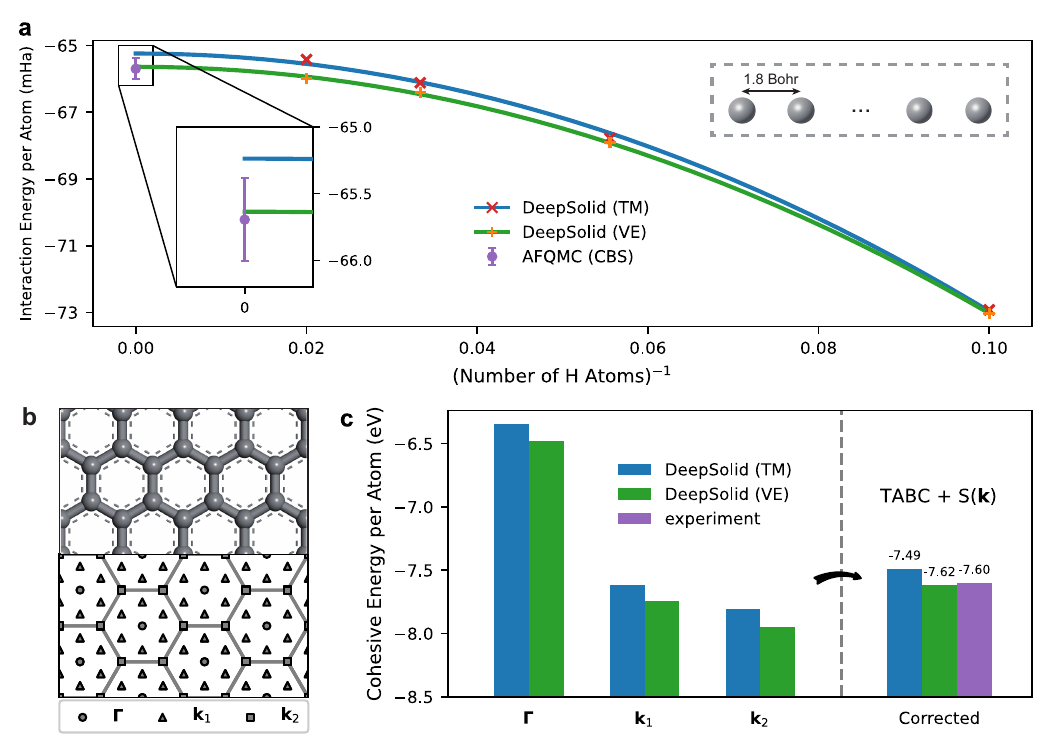}
    \caption{\label{deepsolid}Results of VE method on two periodic systems using DeepSolid~\cite{li2022} as the wavefunction ansatz. \textbf{a}, Interaction energy results of TM and VE methods on the one-dimensional hydrogen chain, whose bond length is fixed at 1.8 Bohr. The curve is fitted based on the energy results of four systems containing increasing numbers of H atoms (respectively 10, 18, 30, 50) to obtain the energy at TDL. The inset zooms into the TDL location, where the benchmark is the AFQMC (CBS) result reported in ref.~\cite{motta2017towards}. \textbf{b}, Top: structure of graphene. Bottom: \textbf{k}-space of graphene, where the positions of three special \textbf{k} points are marked. \textbf{c}, The graphene cohesive energy per atom results at the three special \textbf{k} points. The results labeled ``Corrected'' uses TABC in conjunction with structure factor $\mathrm{S}(\mathbf{k})$ correction to reduce the finite-size error~\cite{li2022,lin2001twist,chiesa2006finite}. The experimental benchmark is obtained from ref.~\cite{dappe2006local}. All the results of TM method are reported by Li et al.~\cite{li2022}, and the variance extrapolation is based on the corresponding original training data.}
\end{figure*}

\begin{figure*}[!htbp]
    \includegraphics[width=160mm]{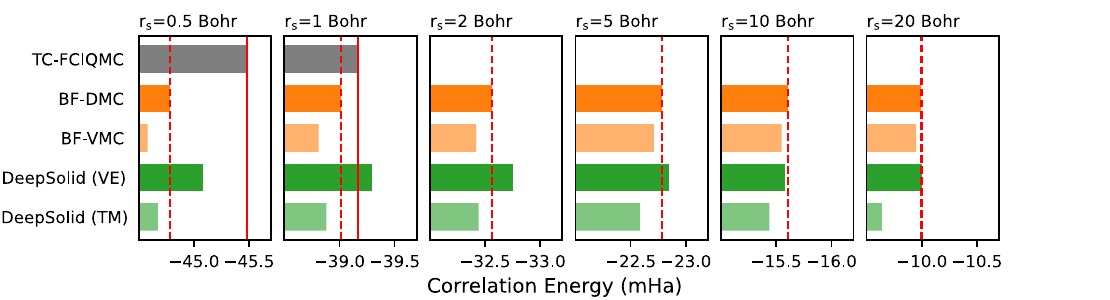}
    \caption{\label{elec_gas}Effects of VE method on homogeneous electron gas using DeepSolid~\cite{li2022} as the wavefunction ansatz. Different panels show the correlation energy results of different densities from $\mathrm{r_s}=0.5$ Bohr to 20.0 Bohr. The BF-DMC, BF-DMC and TC-FCIQMC results are also displayed for comparison~\cite{rios2006inhomogeneous, liao2021towards}. The red line denotes the TC-FCIQMC results as high accuracy benchmark, and the red dashed line points out a upper bound of the exact results due to the variational property of the BF-DMC results. All the results of TM method are reported by Li et al. in DeepSolid paper~\cite{li2022}, and the variance extrapolation is based on the corresponding original training data.}
\end{figure*}

We now discuss the tests of the VM and VE methods on N$_2$ molecule in more detail.
N$_2$ molecule is strongly correlated at the breaking-bond region due to the complexity of triple-bond breaking, providing a challenge case for accurate wavefunction calculation.
Energies at three bond lengths are calculated using FermiNet as the wavefunction ansatz, including $r_0 = 2.0743$ Bohr at the equilibrium geometry, $r_1 = 4.0$ Bohr in the strongly correlated region and $r_2 = 10.0$ Bohr in the completely dissociated region.
As shown in Fig.~\ref{var_matched_extrapolation}a, the energy and variance show linear relationship at all the three bond lengths, which makes it possible for VM and VE methods to be applied.
%

The energy differences relative to the equilibrium geometry $\Delta E_1 = E(r_1) - E(r_0)$ and $\Delta E_2 = E(r_2) - E(r_0)$ are investigated respectively using TM, VM and VE methods, as illustrated in Fig.~\ref{var_matched_extrapolation}b.
%
%
%
Because of the high accuracy of FermiNet on the N$_2$ molecule at bond lengths $r_0$ and $r_2$, the TM method can already provide a good estimate of $\Delta E_2$, leaving only a little room for improvements.
Still and all, it is obvious that the VE method performs better than the VM method.
Regarding $\Delta E_1$, when the accuracy of FermiNet at $r_1$ decreases due to the strong correlation, the TM method becomes inadequate in providing a reliable result.
In this case, the VM method can lead to only a minor improvement, whereas the VE method significantly reduces the error by 67.8\%, from 2.84 mHa to 0.92 mHa.


In Fig.~\ref{var_matched_extrapolation}c, we further investigate the convergence and fluctuation patterns of $\Delta E_1$ and $\Delta E_2$ along the training process when using respectively TM, VM and VE method.
The VE method's curves begin at the training step of $3\times10^4$ in order to ensure a sufficient number of data points for the linear regressions.
Compared with the results obtained through the TM and VM methods, the VE method exhibits faster convergence and a much quicker reduction in fluctuation.
By the $10^4$th training step, the VE method has achieved convergence for both $\Delta E_1$ and $\Delta E_2$, with smooth curves and no fluctuations.
This suggests that using the VE method, VMC training does not require complete convergence, resulting in significant computational cost savings.

To further verify the reliability of the VE method, we tested it on the N$_2$ dissociation curve, which contained calculations of N$_2$ molecule at 25 different bond lengths, involving intact and breaking triple bond (Fig.~\ref{var_matched_extrapolation}d).
For comparison, the state-of-the-art r12-MR-ACPF results~\cite{gdanitz1998accurately} are also displayed.
Considering total energy, all the results show remarkable progress after variance extrapolation, and most results are refined to be within chemical accuracy.
As for the non-parallelity error, i.e. the difference between the maximum and the minimum errors along the curve, the VE method enhances the original result of 4.53 mHa to 2.89 mHa, which is comparable to the r12-MR-ACPF result of 2.14 mHa.
The corresponding extrapolation details are presented in Supplementary Fig.~2.
Since it is a bit arbitrary to choose which of the 25 different bond lengths to match the energy variance, the VM method is not tested here.

Finally, it is worth noting that by improving the accuracy of total energy estimations, it becomes easier to accurately derive certain types of relative energy, such as dissociation energy.
Normally, we have two strategies for calculating the dissociation energy of a system AB containing two parts A and B,
\begin{align}
    \Delta E &= E(A)+E(B)-E(AB),\label{eq6}\\
    \Delta E &= E(A+B) - E(AB),\label{eq7}
\end{align}
where $E(AB)$ denotes the energy of the AB system, $E(A)$ and $E(B)$ denote the energies of the individual monomers A and B, and $E(A+B)$ denotes the energy of the system consisting of separate monomers A and B at a sufficient distance from each other.
If a difference appears in $E(A) + E(B)$ and $E(A+B)$, then it is often regarded as the appearance of size inconsistency of the calculation, which would cause severe problems when computing useful quantities such as the dissociation energy of molecules.
In traditional wavefunction methods the size consistency problem is often induced by an limited ansatz, and in NN-VMC the unbalanced optimization of two different systems is also an important contribution of the problem. 
Curing the relative energy error is effectively improving the size consistency property of NN-VMC calculations.

In Fig.~\ref{var_matched_extrapolation}b, $\Delta E^* = 2E(\mathrm{N}) - E(r_0)$ and $\Delta E_2 = E(r_2) - E(r_0)$ are the dissociation energies of the N$_2$ molecule respectively following Eq.~\ref{eq6} and Eq.~\ref{eq7}, where $E(\mathrm{N})$ denotes the energy of atom N reported by Chakravorty et al.~\cite{chakravorty1993ground}.
Apparently, both $E(r_0)$ and $E(r_2)$ would have upward biases due to the limit of ansatz expressiveness.
Before the extrapolation, $\Delta E_2$ performs better than $\Delta E^*$ as the two biases cancel each other out to a large extent.
Through extrapolation, we obtain an improved estimate of $E(r_0)$ and thus greatly improve the accuracy of $\Delta E^*$ without the requirement of $E(r_2)$, which saves half the computational cost.
Generally speaking, $E(A)$ and $E(B)$ are much simpler to obtain precise results than $E(A+B)$ because of the computational complexity of at least $\mathcal{O}(N^3)$ in terms of the system size for accurate quantum Monte Carlo calculations.
Therefore, the VE method providing better total energy results may help a lot on calculating dissociation energy.

To summarize, we show that the VE approach is an effective method for enhancing the accuracy of total energy and relative energy outcomes without incurring additional computational expenses.
In contrast, it appears that the VM approach may not provide significant benefits for calculations at this level of accuracy.

\subsection{Periodic Systems}

Periodic systems are different from molecules, where the system is in principle infinitely large while molecules contain a finite number of atoms.
Nevertheless, a similar problem would occur because of an improper wavefunction ansatz or unbalanced optimization. 
To compute the useful quantities for periodic systems, such as the cohesive energy of solids, it is required to maintain the size extensiveness of the calculation, otherwise the errors are too significant.
To verify whether the VE method also works for periodic systems and improves the size extensiveness, we further analyze two other calculations on periodic systems reported in DeepSolid~\cite{li2022}, namely the one-dimensional hydrogen chain and the two-dimensional graphene.

Hydrogen chain is a simple but challenging and interesting system.
Fig.~\ref{deepsolid}a shows the interaction energy results of different hydrogen chains containing different atom numbers in a supercell and the corresponding extrapolation result to the thermodynamic limit (TDL).
The extrapolation simply follows the quadratic curve fitting with the symmetry axis fixed at 0.
For comparison, the TDL auxiliary field quantum Monte Carlo (AFQMC) energy result at complete basis set (CBS) limit is also plotted~\cite{motta2017towards}.
Upon observing the magnified results depicted in the inset diagram, it is clear that the TDL energy result utilizing the VE method is more consistent with the AFQMC result.
Returning to the four finite size results, the VE method has led to a downward correction on each of them.
The larger the system, the bigger the VE method's correction, which is both expected and reasonable given that the accuracy of DeepSolid may progressively worsen on larger systems.
The corresponding extrapolation details are shown in Supplementary Fig.~3.

Graphene is a famous two-dimensional material attracting broad attention over last two decades~\cite{geim2011nobel}.
The real space and \textbf{k}-space structure of graphene are illustrated in Fig.~\ref{deepsolid}b.
The positions of three special \textbf{k} points sampled following Monkhorst-Pack mesh~\cite{monkhorst1976special, pack1977special} are also marked, at which the cohesive energy is calculated.
The corrected result is regarded as the weighted average of the results at the three k points under twist-averaged boundary condition (TABC) plus an additional structure factor $\mathrm{S}(\mathbf{k})$~\cite{li2022,lin2001twist,chiesa2006finite}.
Compared with the experimental result, it is obvious that the VE method enhances the result.
The corresponding extrapolation details are displayed in Supplementary Fig.~4.

The homogeneous electron gas (HEG) system presented in DeepSolid~\cite{li2022} is also tested, and the corresponding results are shown in Fig.~\ref{elec_gas}.
In each panel, the correlation energies (i.e., enhancements from the Hatree-Fock method) calculated by various methods are plotted, and a upper bound of the exact result is marked using red dashed lines, which utilizes the variational property of the BF-DMC results~\cite{rios2006inhomogeneous}.
While the initial results prior to extrapolation fall above the upper bound, the results subsequent to extrapolation almost consistently surpass the upper bound.
For the densities ${\rm r_s} = 0.5, 1$ Bohr, where TC-FCIQMC results are available~\cite{liao2021towards}, the red full lines indicate high-accuracy benchmarks. 
The extrapolation results at both densities are closer to the red full lines.
%
%
The corresponding extrapolation details are displayed in Supplementary Fig.~5.

The success of the VE method on the periodic systems further illustrates its universality and reliability.
%

\section{Discussion}
Based on our experience, the VE method should be considered as icing on the cake rather than a universal energy correction scheme.
The premise of this method is small enough energy variance, where the energy results are already very accurate.
For example, in all the calculations we report in this work the energy variance converges to 0.5 Ha$^2$ or less and the energy results are only several millihatrees from the exact results (see Supplementary Fig.~2--5).
The VE method here is to take the last small step forward, from excellence to superexcellence.
However, this small step in total energy may be a big step in relative energy, which makes it essential.
Besides, the slope of energy versus variance linear regression can be a very useful analysis aspect when examining different systems.
Having demonstrated the effectiveness of the VE method, it is worth noting that the applications of VE method to large systems would face two challenges and the method should be executed with care.
%
Firstly, as far as our training can reach, the wavefunction is still far from ground state, and the energy variance remains large.
%
%
Secondly, there is big energy fluctuation during the training process of such large systems, which decreases the data quality and affects the extrapolation results.
%

\section{Conclusion}
In conclusion, we propose a new extrapolation technique based on the training data in NN-VMC, which appears to be an effective and efficient solution for a balanced treatment to obtain better error cancellation when calculating relative energies.
Additionally, this method can surpass the limitation of the wavefunction ansatz and obtain an improved estimation of total energy, which is also important in some cases.
Furthermore, employing the VE method yields quicker convergence of energy results with reduced fluctuations, leading to significant savings in computational costs.
Since this approach has shown great power on both FermiNet and DeepSolid, encompassing both molecular and periodic systems, we expect that it will also greatly contribute to the future of NN-VMC on various powerful ansatz.

\begin{acknowledgments}
The authors thank Xiang Li for sharing data and results. We thank Hang Li for support, Xuelan Wen for relevant literature research, and Liwei Wang, Di He, Chuwei Wang, Du Jiang, Ruichen Li and the rest of the teams for helpful discussions. 
J.C. is funded by the National
Natural Science Foundation of China under Grant No.~92165101 and the Strategic Priority Research Program of Chinese Academy of Sciences under Grant No.~XDB33000000.
\end{acknowledgments}

\appendix


\nocite{*}

\bibliography{ref}

\end{document}


\title{Supplementary Information: Variance extrapolation method for neural-network variational Monte Carlo}

\author[1,2]{Weizhong Fu}
\author[1]{Weiluo Ren}
\author[2,3]{Ji Chen}
\affil[1]{ByteDance Research, Zhonghang Plaza, No. 43,  North 3rd Ring West Road, Haidian District, Beijing, People’s Republic of China}
\affil[2]{School of Physics, Peking University, Beijing 100871, People’s Republic of China}
\affil[3]{Interdisciplinary Institute of Light-Element Quantum Materials, Frontiers
Science Center for Nano-Optoelectronics, Peking University, Beijing 100871, People’s Republic of China}

\maketitle

\section{Technical details in the VE method}
\begin{figure}[!b]
    \centering
    \includegraphics{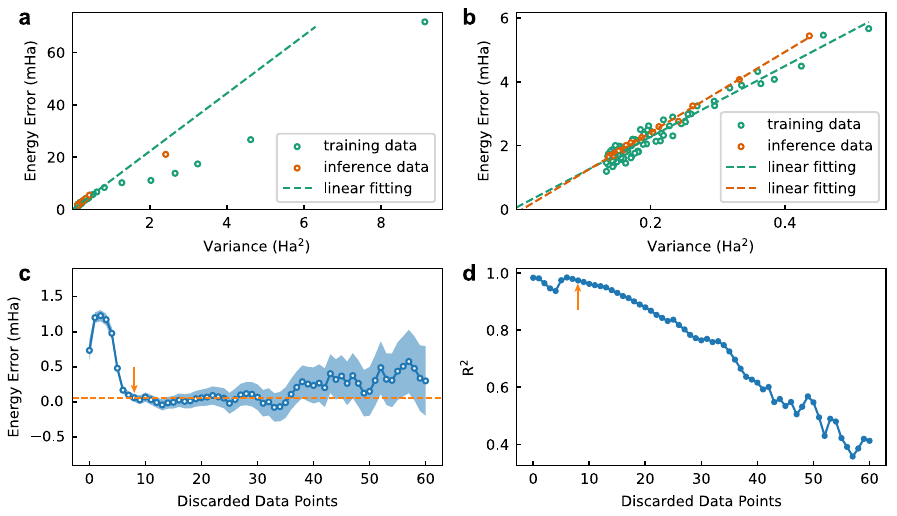}
    \caption{Technical details of VE method on N$_2$ molecule of bond length at $r_0=2.0743$ Bohr. \textbf{a}, The training data and the inference data along the whole training process. The linear fitting result based on the linear portion of the training data is also displayed. \textbf{b}, Zoom into the linear portion of the training data and the inference data. The linear fitting results of both are also displayed. \textbf{c}, Energy error results after extrapolation with different truncation points based on the training data. The shadow denotes the standard errors, i.e. the standard errors of y-intercepts of the linear fittings. The linear fitting shown in the former two panels discards the first 8 data points, which in this panel is indicated by the orange arrow, and the corresponding extrapolation result is shown by the orange dashed line. \textbf{d}, Coefficients of determination $\mathrm{R}^2$ of the linear fittings with different truncation points based on the training data. The orange arrow indicates the same point as in panel \textbf{c}.}
    \label{fig:enter-label}
\end{figure}
We take the calculation on N$_2$ molecule of bond length at $r_0=2.073$ Bohr as an example to demonstrate the technical details in the VE method.
Empirically, we have observed that the linear relationship of energy and energy variance, which is the cornerstone of the VE method, only appears at the near convergence stage.
As shown in Supplementary Fig.~1, in the early training stage, there is a large deviation between the training data and the linear fitting result based on the data close to complete convergence.
It is thus a sensitive issue requiring careful consideration to decide how many data points should we discard from the beginning of the training process.
The extrapolation results and $\mathrm{R}^2$ of the linear fittings with different amounts of data points discarded are displayed in Supplementary Fig.~1c and d, respectively.
The large deviation of the first several data points is reflected in Supplementary Fig.~1c at the beginning where the extrapolation results deviate much from the other results.
After discarding these data points, the extrapolation results become steady and self-consistent, which means there is a good linearity between the left data points.
This is also reflected in Supplementary Fig.~1d on the $\mathrm{R}^2$ curve.
When a set of data has a good linear relationship, deleting the data points on either side of the edge may decrease $\mathrm{R}^2$.
The curve in Supplementary Fig.~1d increases from point number 4 to number 6, which means after discarding the corresponding data point the linear relationship of the remaining data points increases.
After point number 6, $\mathrm{R}^2$ decreases, which means the data points after number 6 exist a good linear relationship.
Finally, we choose to discard 8 data points, whose extrapolation result is within a good agreement with the other extrapolation results considering the linear fitting error.
Note that the standard errors shown in Supplementary Fig.~1c only characterize the uncertainty of y-intercepts caused by linear fittings.
They contain no valid information about the bias from the exact value.
Therefore, we choose to not show error bars for extrapolation results in the main text to avoid confusion.

We also display the inference results at every $10^4$ training steps in Supplementary Fig.~1a and b.
During the early training stage, there is a significant disparity between the training and the inference data, indicating that the training data is heavily skewed at this time.
This is another reason why we should not use these data points when doing linear extrapolation.
The linear fitting of the inference data is also presented in Supplementary Fig.~1b, with a deviation of only 0.17 mHa between the corresponding extrapolated value and the training data result, a negligible difference.
Therefore, all the extrapolations presented in the main text rely solely on the training data rather than inference data to save computing costs.

\section{Extrapolation details}

\begin{figure}[!b]
    \centering
    \includegraphics[width=155mm]{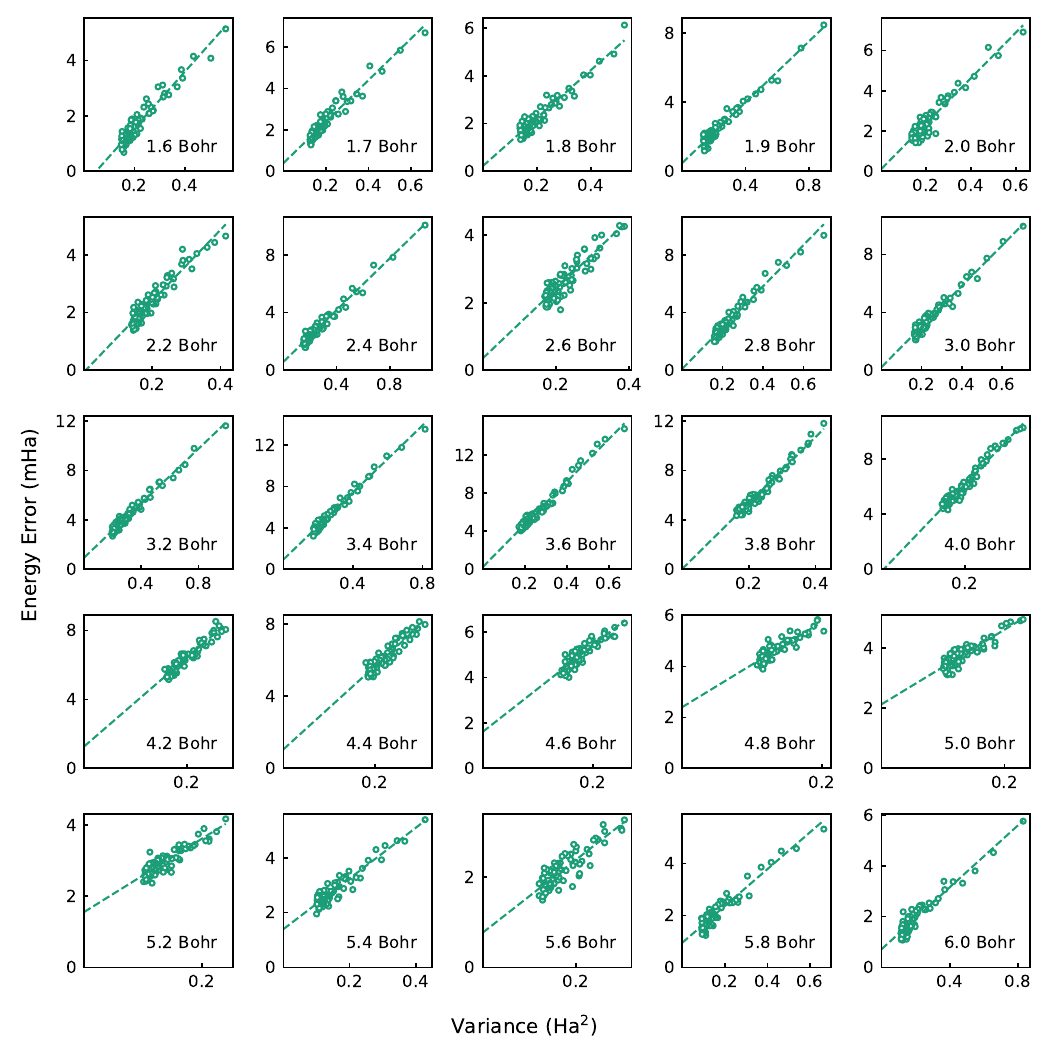}
    \caption{The linear extrapolation results of N$_2$ molecules at different bond lengths from 1.6 Bohr to 6.0 Bohr.}
    \label{fig:enter-label}
\end{figure}

Here we display the extrapolation details including selections of the truncation and results of the linear fitting of all the calculations we mentioned in the main text, respectively the N$_2$ molecules in Supplementary Fig.~2, the hydrogen chain systems in Supplementary Fig.~3, the graphene systems in Supplementary Fig.~4 and the electron gas systems in Supplementary Fig.~5. There is a good linear relationship between the energy and the variance in almost all the systems.

\begin{figure}
    \centering
    \includegraphics{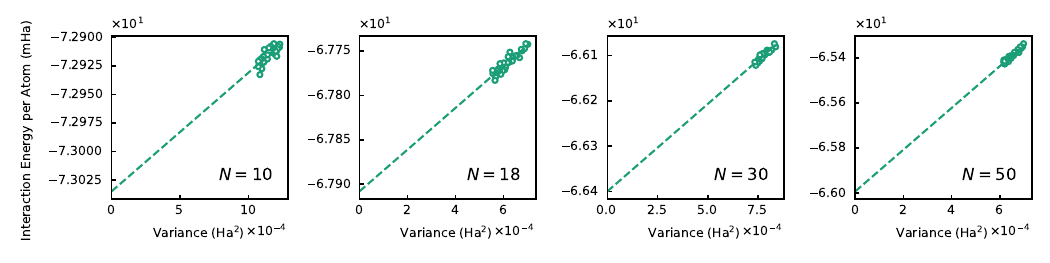}
    \caption{The linear extrapolation results of hydrogen chain systems containing different numbers of hydrogen atoms from $N=10$ to $N=50$.}
    \label{fig:enter-label}
\end{figure}

\begin{figure}
    \centering
    \includegraphics{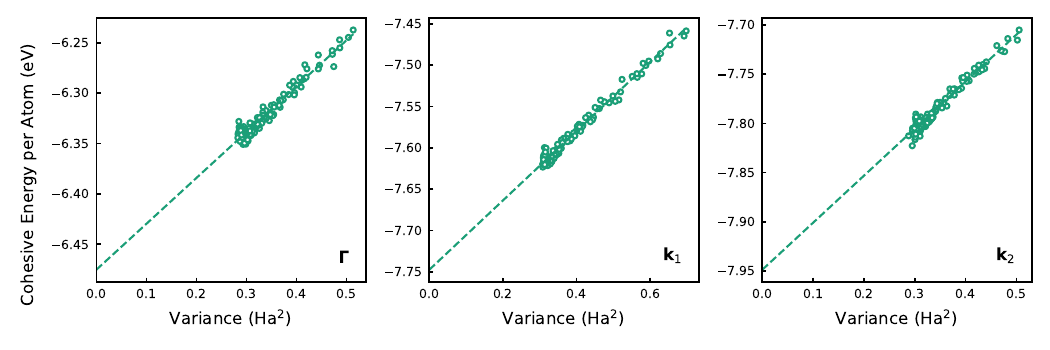}
    \caption{The linear extrapolation results of graphene systems at different special $\mathbf{k}$ points, respectively $\mathbf{\Gamma}$, $\mathbf{k}_1$ and $\mathbf{k}_2$.}
    \label{fig:enter-label}
\end{figure}

\begin{figure}
    \centering
    \includegraphics{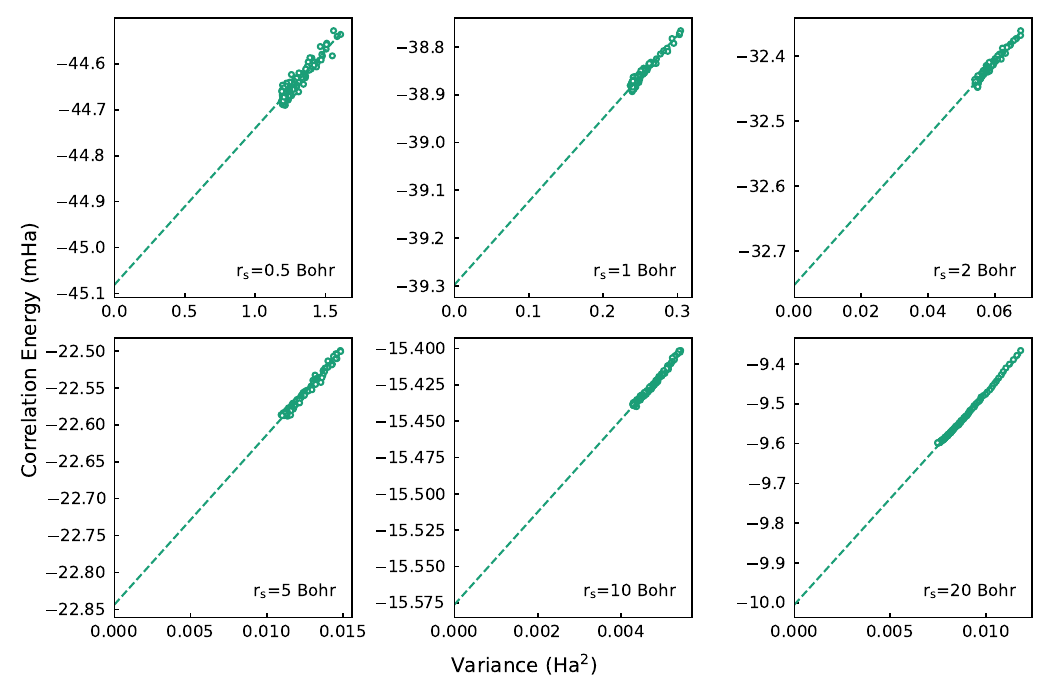}
    \caption{The linear extrapolation results of electron gas systems at different densities from $\mathrm{r}_s=0.5$ Bohr to $\mathrm{r}_s=20$ Bohr.}
    \label{fig:enter-label}
\end{figure}